\documentclass{aa}

\usepackage{graphicx}

\begin{document}

\title{Nova V5116 Sgr and searching for superhumps in nova remnants}

\author{A.~Dobrotka \inst {1}, A.~Retter \inst {2}, A.~Liu \inst {3}}

\offprints{A.~Dobrotka}

\institute{Departement of Physics,
           Faculty of Materials Science and Technology,
           Slovak University of Technology in Bratislava,
           J\'ana Bottu 25,
           91774 Trnava,
           The Slovak Republic,
           (e-mail:~andrej.dobrotka@stuba.sk)
\and
           Departement of Astronomy and Astrophysics,
           Penn State University,
           525 Davey Lab, University Park,
           PA, 16802-6305,
           USA
           (e-mail:~aretter@walla.com)
\and       
           Norcape Observatory,
           PO Box 300,
           Exmounth, 6707,
           Australia
	   (asliu@bigpond.net.au)
           }

\date{Received / Accepted}

\abstract
   {}
   {We present the period analysis of unfiltered photometric observations of
     V5116~Sgr (Nova Sgr 2005 \#2) and we search for superhump candidates in
     novae remnants.}
   {The PDM method for period analysis is used. The masses of the novae
     componets are estimated from the secondary mass -- orbital period and
     primary mass -- decline time relations.}
   {We found that 13 nights of V5116~Sgr observations in the year 2006 are
     modulated with a period of $0.1238 \pm 0.0001$~d ($2.9712 \pm
     0.0024$~h). Following the shape of the phased light curves and no
     apparent change in the value of the periodicity in different subsamples
     of the data, we interpret the period as orbital in nature. The binary
     system then falls within the period gap of the orbital period
     distribution of cataclysmic variables. From the maximum magnitude -- rate
     of decline relation, we estimate the maximum absolute visual magnitude of
     $M_{\rm Vmax} =  -8.85 \pm 0.04$~mag using the measured value of
     decline $t_{\rm 2} = 6.5 \pm 1.0$~d. The mass-period relation for
     cataclysmic variables yields a secondary mass estimate of about $0.26 \pm
     0.05~{\rm M}_{\rm \odot}$. We propose that V5116~Sgr is a high
     inclination system showing an irradiation effect of the secondary
     star. No fully developed accretion disc up to the tidal radius with the
     value lower than $3.5~10^{10}$~cm is probable. The mass ratio was
     estimated in a few novae and the presence or absence of superhumps in
     these systems was compared with the mass ratio limit for superhumps of
     about 0.35. We found that in the majority of novae with expected
     superhumps, this variability has not been found yet. Therefore, more
     observations of these systems is encouraged.}
   {}

\maketitle \markboth{A.~Dobrotka et al.: Nova V5116 Sgr and searching for
  superhumps in nova remnants.}{}

\section{Introduction}
%\label{}

Novae are a subclass of cataclysmic variable stars. In these interracting
binaries, the white dwarf is accreting the matter transfered from the
secondary star. The accretion disc may be formed in the non magnetic case. The
intermediate polar systems have a truncated disc and polar systems have
magnetic field strong enough to prevent the disc formation (see Warner 1995
for review). The accumulation of critical amount of accreted material onto the
white dwarf surface results in a nova explosion. The distribution of orbital
periods in cataclysmic variables shows a period gap between about 2 and 3
hours. Novae do not show this lack of objects in the mentioned interval.

V5116~Sgr (Nova Sgr 2005 \#2) was discovered by Liller (2005) on 2005 July
4.049 UT. The nova had a magnitude $\sim 8.0$ on two red photographs. An
unfiltered CCD image from 2005 July 5.085 UT showed the object at mag 7.2. The
spectrum from 2005 July 5.099 UT showed H$_{\rm \alpha}$ with the FWHM of
$\sim 970$ km~s$^{-1}$. The expansion velocity derived from the sharp P~Cyg
profile was $\sim 1300$ km~s$^{-1}$. The position of the nova was measured by
Gilmore and Kilmartin (2005) and Jacques (2005). Gilmore and Kilmartin (2005)
searched for the nova precursor, but no convincing candidate has been
found. Russell et al. (2005) performed a 0.8 -- 2.5 $\mu$m spectroscopy of the
nova on 2005 July 15. The object showed emission lines of H~I, He~I, C~I, N~I,
Ca~II and O~I with a FWHM $\sim$ 2200 km~s$^{-1}$. He~I showed P~Cyg profile
at 1.0830 and 2.0581 $\mu$m. No thermal dust emission was observed.

After the nova explosion the accretions disc is destroyed. The new disc is
reformed by the stream of matter flowing from the secondary, interracting with
itself and forming a ring in the circularisation radius. The disc starts to
form by the viscous shearing. Matter losing angular momentum is moving invard
and the excess of angular momentum is transported by the matter flowing
outward. The invard moving matter touchs the white dwarf and the disc is
reformed (Pringle 1981). If the white dwarf is magnetic, the matter interracts
with the magnetosphere and is then conducted by the magnetic field to the
magnetic poles. The interraction of the flow with the poles of the rotating
star is observed as periodic signal with the spin period of the white
dwarf. In the case of intermediate polars the spin period is usually much
shorter than the orbital period (Patterson 1994, Hellier 1996). The polar
systems without a disc are synchronous rotators, hence the spin period is
equal or close to the orbital period (see e.g. Schmidt and Stockman
1991). Nova V1500 Cyg (polar system without a disc) changed the period from
0.141~d to 0.138~d and then stabilised at 0.140~d (Patterson 1978, 1979). The
difference of 1.8\% between the rotation period of the white dwarf and the
orbital period is ascribed to the effects of the nova explosion in 1975. The
synchronisation can be then corrupted by the nova explosion, but the spin
period remain still very close to the orbital period. This is in contrast with
intermediate polars with disc presence.

Searching for periods in novae allows to study the orbital distrubution and
evolution of these systems. Currently, there are about 50 novae with known
orbital periods (Warner 2002) with typical values ranging from 2 to 9
hours. The existence of the accretion disc or its renovation after the nova
explosion is confirmed by the detection of the superhump period (see
e.g. Retter et al. 1997, Kang et al. 2006a) or the spin period of a magnetic
white dwarf in the case of intermediate polars (see e.g. Retter et al. 1998).

Superhumps are caused by a precessing accretion disc generally in systems with
mass ratio $M_{\rm 2}/M_{\rm 1} < 0.35 \pm 0.02$ (see Patterson et al. 2005
for review), in which the disc radius reaches the 3:1 resonance. The mass
ratio indicates that systems with massive primaries and low mass secondaries
are probable superhumpers. Novae in general possess high mass white dwarfs
(see e.g. Warner 1995, Smith and Dhillon 1998, Webbink 1990) and short orbital
periods sugest low mass secondaries. It is therefore meaningfull to search for
superhump variability in novae with short orbital period.

We have an ongoing program to observe novae with small telescopes to search
for periodicities in their optical light curves. In this paper we report the
detection of one periodicity in our photometric data ($P = 0.1238 \pm
0.0001$~d) of V5116~Sgr and we discuss the presence of superhumps in nova
remnants. In Section~\ref{obs} we present our observational material. The
long-term light curve with the period analysis of the data is presented in
Sec.~\ref{data_anal} and in Sec.~\ref{disc} we discuss the long-term light
curve behaviour ({Sec.~\ref{disc_v5116_l}}), the results of the period analysis
(Sec.~\ref{disc_v5116_s}) and superhump search in nova remnants
(Sec.~\ref{disc_SH}).

\section{Observations}
\label{obs}

V5116 Sgr was observed during 13 nights in 2006. The observations included
45.2 hours and 1256 data points in total. The list of observations is
presented in Table~\ref{obslog}. The photometry was carried out with a 0.3-m
f/6.3 telescope coupled to a SBIG ST7E CCD camera. The telescope is located in
Exmounth, Western Australia. The pixel size of the CCD is 9 x 9 microns. This
camera is attached to an Optec f5 focal reducer giving an image field of view
of 15 x 10 arcmin. The range of seeing for the data was 2.5 -- 3 arcsec. The
exposure times were between 30 and 90 sec every 120 sec, and no filter was
used. Aperture photometry was used in the reduction, with an aperture size of
12 pixels (radius). As comparison star we used GSC 7392 856 with V = 12
  mag and as check star GSC 7392 292 with V = 11.6 mag. The instrumental
  magnitudes of the stars were derived from the SBIG CCDOPS software and the
  magnitudes of the comparison and check stars from the GSC.

\begin{figure}
\includegraphics[width=90mm]{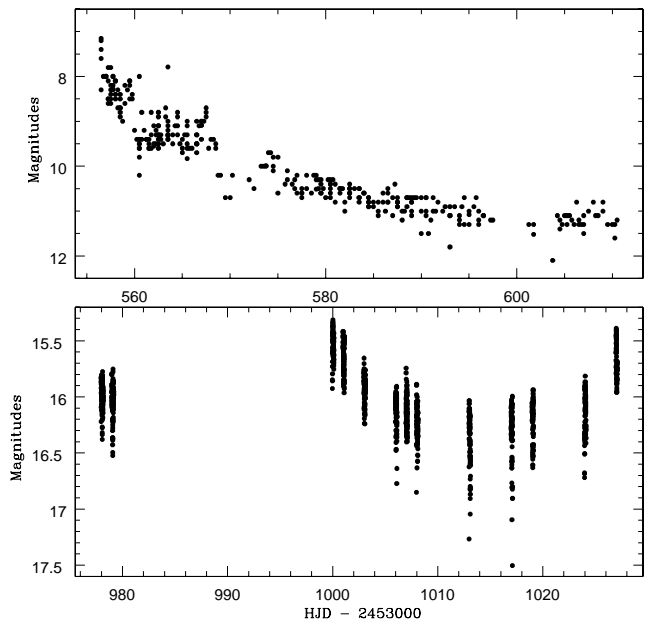}
\caption{The AAVSO light curve (upper panel) and the unfiltered light curve of
  our data (bottom panel).}
\label{curve}
\end{figure}

\begin{figure}
\includegraphics[width=90mm]{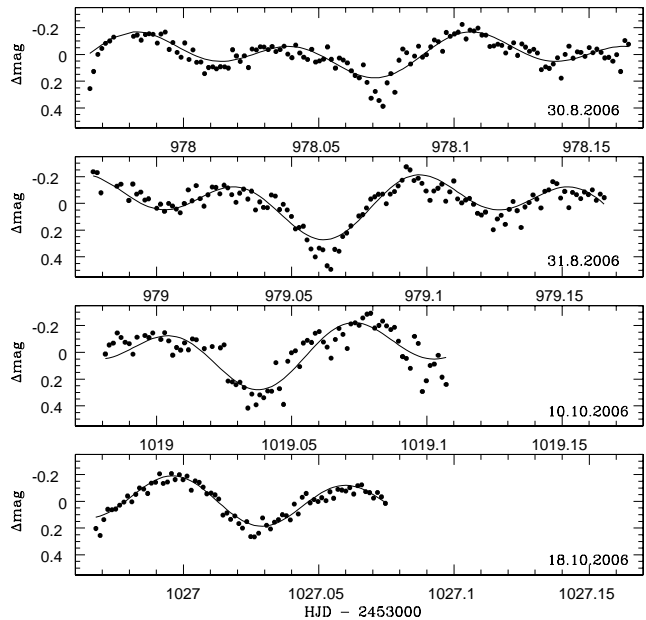}
\caption{A selection of our observations (4 detrended best runs; 1,2,11 and
  13). The solid curve is the sinusoidal fit (plus first harmonic) to the data
  using the period derived in this paper.}
\label{4runs}
\end{figure}

Figure~\ref{curve} displays our light curve and the long-term light curve of
the nova from the American Association of Variable Star Observers (AAVSO). In
Fig.~\ref{4runs} we show 4 best examples of our observational runs with the
sinusoidal fit (with first harmonic) to the data using the period derived in
this paper. The studied variations are clearly seen.
\begin{table}
\caption{The observational log. HJD is --2453000 days, $t$ is the duration of
  the observation in hours and $N$ is the number of frames.}
\begin{center}
\begin{tabular}{lcccr}
\hline
\hline
 Run & Date & HJD (start) & $t$ & $N$ \\
 & & & [h] & \\
\hline
1 & 2006 August 30 & 977.9654 & 4.8 & 133\\
2 & 2006 August 31 & 978.9766 & 4.5 & 123\\
3 & 2006 September 21 & 999.9753 & 3.4 & 94\\
4 & 2006 September 22 & 1000.9831 & 3.3 & 93\\
5 & 2006 September 24 & 1002.9704 & 3.6 & 101\\
6 & 2006 September 27 & 1005.9734 & 3.5 & 93\\
7 & 2006 September 28 & 1006.9792 & 3.3 & 90\\
8 & 2006 September 29 & 1007.977 & 3.4 & 98\\
9 & 2006 October 04 & 1012.974 & 3.4 & 94\\
10 & 2006 October 08 & 1016.9812 & 3.2 & 90\\
11 & 2006 October 10 & 1018.9811 & 3.0 & 82\\
12 & 2006 October 15 & 1023.9679 & 3.2 & 91\\
13 & 2006 October 18 & 1026.9677 & 2.6 & 74\\
\hline
\hline
\end{tabular}
\end{center}
\label{obslog}
\end{table}

\section{Data analysis}
\label{data_anal}

\subsection{Long-term light curve}
\label{t2t3}

By using the maximum visual magnitude of $V =7.15$ we measured the time
required for a decline of two magnitudes from maximum $t_{\rm 2} = 6.5 \pm
1.0$~d and three magnitude from maximum $t_{\rm 3} = 20.2 \pm 1.9$~d
(Fig.~\ref{curve} -- upper panel). This makes V5116~Sgr a very fast nova
according to the classification given in Table 5.4 of Warner (1995). We
calculated the visual absolute magnitude at maximum brightness using Equation
(2) of Della Valle and Livio (1995) and obtained $M_{\rm Vmax} = -8.85 \pm
  0.04$ mag.

The bottom panel of the Fig.~\ref{curve} shows our unfiltered photometry of
the nova. The long-term variations with time scale $\sim 30$ -- $40$ days with
amplitude $\sim 1$~mag $\sim 450$ days after the outburst is
clear. Similar large scale quasi-periodic brightness oscillations with
amplitude 1 -- 1.5~mag months after the maximum were observed for example in
GK~Per, V603~aql, DK~Lac (see e.g. Warner 1995) or V4745~Sgr (Cs\'ak et
al. 2005), with time scales 5,12,25 and $\sim 20$ days respectively.

\subsection{Period analysis}

For the period analysis we used the PDM method (Stellingwerf 1978). First, the
data were detrended by substracting the linear fit from each night. A simple
visual inspection of the runs suggests that the data are modulated with a
periodicity of $\sim 1.5$ hours. This value was then identified as the first
harmonic of the real periodicity, therefore the exact value of the frequency
is $f = 8.080  \pm 0.006$~d$^{-1}$, which corresponds to the periodicity $P =
0.1238 \pm 0.0001$~d ($2.9712 \pm 0.0024$~h). We calculated the errors in the
frequency from the half width at the half minimum of the peaks in the PDM
power spectrum.

\begin{figure}
\includegraphics[width=90mm]{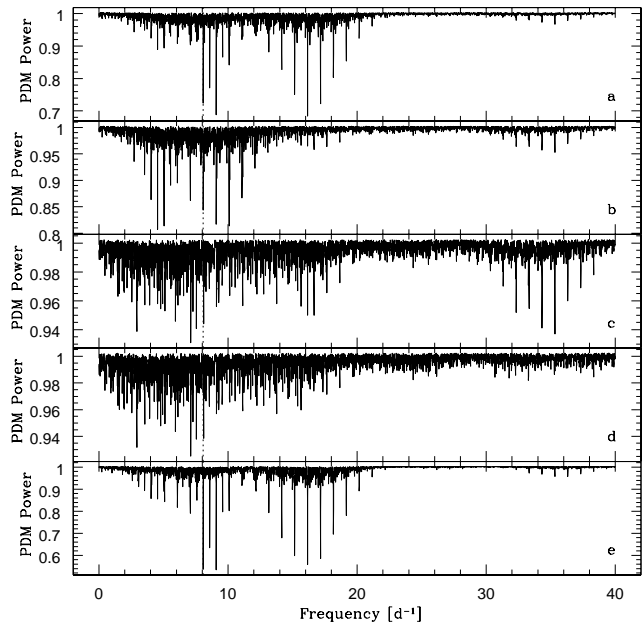}
\caption{Power spectra of all the data. Panel a -- raw detrended data (night
  means substracted), panel b -- data after the substraction of the harmonic
  frequency $2 \times f$, panel c -- data after the substraction of the
  harmonic $2 \times f$ and the fundamental frequency $f$, panel d -- data
  after the substraction of the harmonic $2 \times f$, the fundamental
  frequency $f$ and the harmonic frequency $4 \times f$, panel e -- synthetic
  spectrum with the frequencies $f$, $2 \times f$ and $4 \times f$. The dashed
  line represent $f = 8.08$~d$^{-1}$.}
\label{power}
\end{figure}

\begin{figure}
\includegraphics[width=90mm]{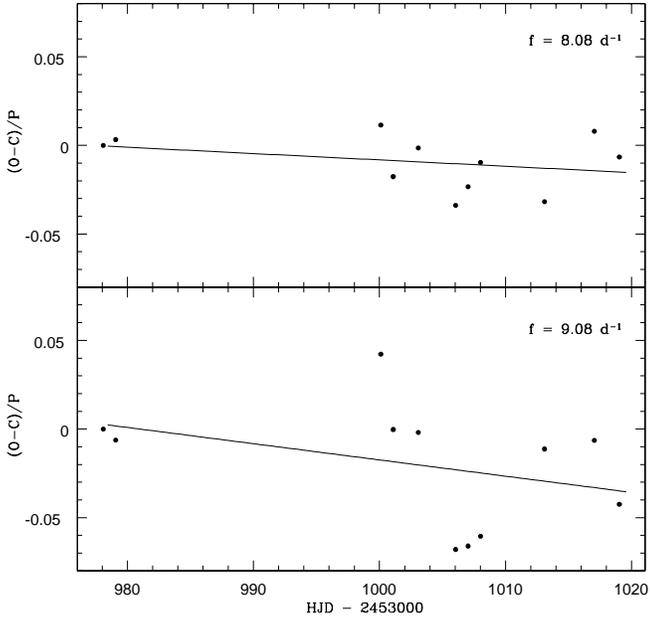}
\caption{O--C diagram. Upper panel -- for frequency $8.080 \pm 0.006$~d$^{-1}$
  ($P = 0.1238 \pm 0.0001$~d) and lower panel -- for frequency $9.080 \pm
  0.006$~d$^{-1}$ ($P = 0.1101 \pm 0.0001$~d). The solid line is the linear
  fit to the data.}
\label{o-c}
\end{figure}

\begin{figure}
\includegraphics[width=90mm]{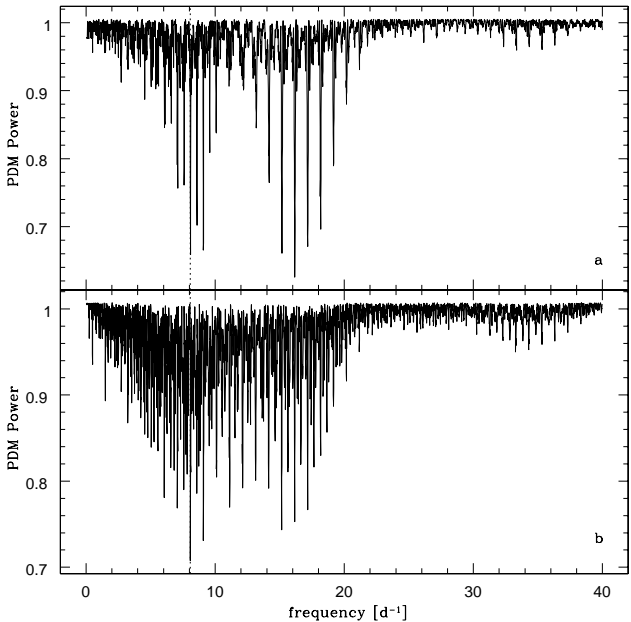}
\caption{Power spectra of two different subsamples of the data. Upper panel --
  runs 3 -- 8; lower panel -- runs 9 -- 13. The dashed line represents $f =
  8.08$~d$^{-1}$.}
\label{power2}
\end{figure}

The results from the period analysis of all the data are depicted in
Fig.~\ref{power}. The suspected periodicity ($f = 8.080 \pm 0.006$~d$^{-1}$,
$P = 0.1238 \pm 0.0001$~d) is seen with its 1-d aliases and the dominant peak
is the first harmonic ($2 \times f = 16.160 \pm 0.006$~d$^{-1}$). For testing
which value is the real signal and which is the alias we performed an O--C
diagram with a linear fit to the data (Fig.~\ref{o-c}; upper panel -- O--C for
the frequency $8.08$~d$^{-1}$ and lower panel -- O--C for the frequency
$9.08$~d$^{-1}$). The lower panel shows a larger scatter of the O--C values
compared with the upper panel. This indicates that the real signal is $f =
8.080 \pm 0.006$~d$^{-1}$ and the dominant peak at $\sim 16$~d$^{-1}$ is the
first harmonic of the real signal. To test the reality of the peaks we
performed a power spectrum of the synthetic data using a sinusoids with the
derived periods ( including fundamental $f$, first harmonic $2 \times f$
and $4 \times f$ described later) and phases with amplitudes found from
the fits to the data sampled as the original data with introduced Gaussian
noise (Fig.~\ref{power}e). The similarity with the original periodogram is
clear. Further we performed a period analysis of two different subsamples of
the data (first subsample -- runs 3 -- 8; second subsample -- runs 9 -- 13)
depicted in Fig.~\ref{power2}. The frequency is present together with its
first harmonic in both datasets. The values are $8.090 \pm 0.015$~d$^{-1}$ and
$8.080 \pm 0.028$~d$^{-1}$ (first and second subsample respectively) which is
within the errors in both cases. We also performed an analysis of the check
star and did not identify any periodicity associated with its data.

There is a strong peak near the value $\sim 8.58$~d$^{-1}$. This signal is not
real as seen in Fig.~\ref{power}b,c. After the substraction of $2 \times f$
(Fig.~\ref{power}b) this peak dissapeared and after a further substraction of
$f$ (Fig.~\ref{power}c) no peak is dominant in this frequency region. This
feature also appears in the synthetic power spectrum (Fig.~\ref{power}e), so
it is definitely an alias of $f$.

Another strong feature is a peak at $\sim 35$~d$^{-1}$. Substracting the
harmonic frequency $4 \times f$ from the data, after the substraction
described in the previous paragraph, this signal dissapared
(Fig.~\ref{power}d). It is another artefact of the real signal $f$. No other
dominant peaks are visible with frequencies higher than $40$~d$^{-1}$. No
other frequency was then found.


\subsection{The structure of the periodicity}

\begin{figure}
\includegraphics[width=90mm]{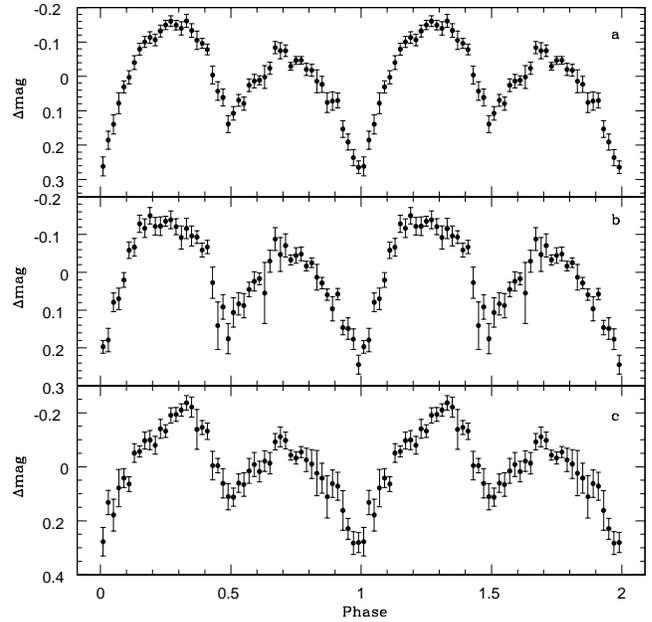}
\caption{The light curve of V5116~Sgr folded on the $0.1238$~d period and
  binned into 50 equal bins. Panel a -- all the data, panel b -- runs 3 -- 8;
  panel c -- runs 9 -- 13.}
\label{folded}
\end{figure}

In Figure~\ref{folded} we show the light curve of V5116~Sgr folded on the
$0.1238$~d period. The points are averaged magnitudes in 50 equal bins that
cover the 0 -- 1 phase interval. The full amplitude of the mean variation is
$0.43 \pm 0.02$~mag for primary minimum and $0.30 \pm 0.02$~mag for secondary
minimum. Using the sinusoidal fit to determine the amplitude is not adequate,
because of the non sinusoidal shape of the light curve, we therefore estimated
the amplitude by measuring the difference between extrema. The bars are the
errors of the mean value. The light curves folded on the periodicity found
show typical features of an eclipsing system.

We determided the epoch of minima by fitting low-order polynomials to selected
parts of the light curve. The best fitted (using sinusoidal fits) ephemeris of
the periodicity is:
\vskip2mm
$T_{\rm (min)}(HJD) = 2453978.0726 + 0.1238 E$\\
$~~~~~~~~~~~~~~~~~~~~~~~~~~~~~~~~\pm 0.0009$~$\pm 0.0001$

\section{Discussion}
\label{disc}

\subsection{Long-term variations of V5116~Sgr}
\label{disc_v5116_l}

The long-term light curve of the nova V5116~Sgr (Fig.~\ref{curve} --
  bottom panel) shows a transition from smooth decline to probable
  oscillations. Several models has been suggested for such
  ``transient'' phase. One of them explains the transition as the time when
the accretion disc is re-established (Retter 2002). The author propose a
connection between this phase and intermediate polars. There is no indication
of magnetic nature of the white dwarf in V5116~Sgr yet. Spin period of the
white dwarf indicating the intermediate polar type of this ``transient''
  phase in novae was detected for example in V4745~Sgr by optical
observations (Dobrotka et al. 2006a), V1494~Aql (Drake et al. 2003) and
V4743~Sgr by X-ray observations (Ness et al. 2003). If this
  ``transient'' phase interpretation of Retter (2002) is applicable in the
case of V5116~Sgr, the accretion must be then re-established and the accretion
disc should be present.

\subsection{Short-term variations of V5116~Sgr}
\label{disc_v5116_s}

We have identified one periodicity in the light curve of the nova V5116~Sgr
about 15 months after the maximum brightness. The value is $P = 0.1238 \pm
0.0001$~d ($2.9712 \pm 0.0024$~h). The upper limits of the period
difference between two subset of data with $\simeq 16$ days of mean distance
analysed in Fig.~\ref{power2} is $1.5~10^{-4}$~d which gives $|\dot{P}| \simeq
0.94~10^{-5}$. Following Patterson et al. (1993) the mean variation of the
superhump period in SU~UMa superhumping sytems is $|\dot{P}| \simeq
3-10~10^{-5}$ (see their Table.~1). For the recurrent nova VY~Aqr the authors
derived a variation of $|\dot{P}| \simeq 8.2~10^{-5}$. Our value of
  $|\dot{P}|$ comes probably from the errors of period measurements rather
  than from real period changes. The period seems then to be constant
during our observations and the shape of the folded light curve suggests a
primary and a secondary eclipse. We therefore propose that this periodicity is
the orbital period of the binary system. Such a period is at the lower end of
the mostly populated region of orbital periods in novae (Warner 2002). The
dominant first harmonic frequency in the power spectra is a result of the
clear structure primary -- secondary eclipse in the folded light curve which
suggests a high inclination angle of the binary system. Using the orbital
period of the system and equation (9) from Smith and Dhillon (1998) we obtain
a rough estimate for the secondary star mass of $0.26 \pm 0.05$~M$_{\rm
  \odot}$. Using a mean white dwarf mass of $0.85 \pm 0.05$~M$_{\rm \odot}$
from Smith and Dhillon (1998), we find a mass ratio (secondary/primary mass)
of $M_{\rm 2}/M_{\rm 1} = 0.3 \pm 0.1$.

After the nova explosion, the hot white dwarf may heat and irradiate the
cooler companion. The observed orbital light curve of the nova can
be the result of the aspect variations or eclipses of the secondary due to
heating from the hot primary and the asymmetry in the pulse profiles could be
produced once the shape of the secondary is of a tear drop model. The
irradiation effects in classical novae can also be detected long after the
outburst stage (e.g., V1500~Cyg; Sommers and Naylor 1999, DN~Gem; Retter et
al. 1999). Two eclipse like features are present in the folded light curve of
V5116~Sgr (Fig.~\ref{folded}). The shape is similar to the light curve of
V2540~Oph in 2003 (Ak et al. 2005) with a dip at phase $\sim 0.5$ ($\sim 33\%$
amplitude of the primary minimum for V2540~Oph and $\sim 70\%$ for
V5116~Sgr). The authors concluded that V2540~Oph is likely a high inclination
system either showing an irradiation effect or having a spiral structure in
its accretion disc. Woudt and Warner (2003a) noted that one of the following
requirements must be fulfilled for the large amplitude orbital modulations to
be seen in the light curve of a recent nova in which the accretion disc does
not dominate the luminosity of the system: 1) the disc is foreshortened but
the secondary is seen (a high inclination angle), 2) the disc has small
dimensions, 3) no disc (case of polars). In the case of V5116~Sgr the polar
interpretation is possible because of the orbital period distribution of
polars which peaks below 5 hours (Warner 1995), but nothing else indicate this
option. The small dimension of the disc is supported by; 1) the short orbital
period, 2) the post nova stage when the disc is reforming after explosion.

A comparison to two novae within the period gap IM~Nor and DD~Cir with present
irradiation effect (Woudt and Warner 2003b,c) can be made. The light curve of
V5116~Sgr is different from those of IM~Nor and DD~Cir but similar to
V2540~Oph. IM~Nor and DD~Cir showed very small dips at phase 0.5 interpretted
as partial eclipses of the irradiated secondary by the disc or matter
stream. The light curve shape depends on the disc radius and on the
inclination angle. The deep secondary eclipses in V5116~Sgr require a large
disc or a high inclination. The strength of the irradiation effect (white
dwarf post-outburst temperature) can also play a significant role. The
differences in the phase of the maxima in Fig.~\ref{folded}b,c can be due to
spiral structures in the disc as mentioned before in regard with the nova
V2540~Oph.

The reconstruction of an accretion disc after the nova eruption is indicated
by the discovery of the white dwarf spin or by the superhump period. A
probable spin period of the white dwarf was detected $\sim 1$ year after the
outburst in V4745~Sgr (Dobrotka et al. 2006a), $\sim 2.75$ years in V4743~Sgr
(Kang et al. 2006b) and $\sim 15$ months after the maximum in V1425~Aql
(Retter et al. 1998). Several systems show superhumps as early as two and a
half months after the outburst like V4633~Sgr (a spin period is another option
in this case, Lipkin et al. 2001) or two years after the outburst like
V1974~Cyg (Retter et al. 1997). In our V5116~Sgr light curves extending
  $\sim 470$ days after outburst we did not find any photometric variations
  consistent with white dwarf spin modulation or superhump properties. We
can not say anything about the magnetic nature of the white dwarf, but the
components mass ratio indicates that the superhump existence is probable. The
mass ratio using the primary mass average is not decisive. Using the derived
secondary mass 0.26~M$_{\rm \odot}$ (Section~\ref{disc_v5116_s}) the mass
ratio is lower than 0.35 for primary mass higher than $0.74$~M$_{\rm
  \odot}$. Taking all known primary masses in cataclysmic variables (Ritter
and Kolb 2003), 63\% have higher or equal mass than $0.74$~M$_{\rm
  \odot}$. This probability value is not enought to make sure that superhumps
are expected inV5116~Sgr, but the search for such variability could be
fruitful. However V5116~Sgr is a very fast nova. According to equation (13)
from Livio (1992) and taking $t_{\rm 3} = 20.2 \pm 1.9$~d derived in this
paper (Section~\ref{t2t3}) we obtain a mass estimate of the white dwarf of
$1.04 \pm 0.02$~M$_{\rm \odot}$ and thus $M_{\rm 2}/M_{\rm 1} = 0.25 \pm
0.05$. The presence of superhumps is then expected if a disc is fully
developed up to tidal radius.

The superhump period is a few percent longer or shorter than the orbital
period. The possibility that the periodicity found in this paper is a
superhump is rejected following the stability discussion and the shape
of the folded light curve (Fig.~\ref{folded}). The mean shape of
superhumps is typically an asymetric sinusoid and our data show typical
eclipse like features.

The presence of the disc in cataclysmic variables depends on the mass loss
from the secondary. The matterial supplied from the secondary within the
period gap depends on the strength of the magnetic braking driving the
secondary out of thermal equilibrium (mass loss time scale shorter than the
thermal time scale). The stronger the braking, the wider the gap will be and
the higher is the upper end of the period gap. If the magnetic braking is low
enought, the mass loss time scale may never become shorter than the thermal
time scale. In the case of novae another condition is important. The nova
explosion heats the secondary which leads to an enhanced mass
transfer. Therefore the complete absence of accretion discs in novae within
the period gap is not expected.

\subsection{Searching for superhumps in other nova remnants}
\label{disc_SH}

The orbital period 2.462~h of the mentionned IM~Nor (Woudt and
Warner 2003b) yields a secondary mass estimate
of 0.20~M$_{\rm \odot}$ following Smith and Dhillon (1998). The decay time
$t_{\rm 3} \simeq 50$~d (Kato et al. 2002) yields a primary mass estimate of
$\simeq 0.86$~M$_{\rm \odot}$ (Livio 1992) giving a mass ratio $\simeq
0.23$. Woudt and Warner (2003b) concluded that they did not observe the
superhump modulation. This conclusion with the mass ratio safely lower than
0.35 indicate that the existence of a fully developed disc is not
probable. DD~Cir (Woudt and Warner 2003c) has a period of 2.339~h yielding
a secondary mass estimate of 0.18~M$_{\rm \odot}$. The decay time $t_{\rm 2}
\simeq 4.5$~d (Liller 1999, no other estimates are available) place the object
in the class of very fast novae (similar to V5116~Sgr). The decay time $t_{\rm
  3} \simeq 10$~d following the equation $t_{\rm 3} \simeq 2.75~t_{\rm
  2}^{0.88}$ from Warner (1995) gives a rought white dwarf mass estimate of
$\simeq 1.16$~M$_{\rm \odot}$ (mass ratio $< 0.2$). Woudt and Warner(2003b)
interpretted the deep eclipse as obscuration of the disc and did not detect
any periodicity near the orbital period in the period analysis. Therefore,
following the tidal instability model it is again probable that the disc is
not fully developed up to 3:1 resonance radius 3 years after the
outburst. However the authors argued that the disc radius is 47 \% of the
orbital separation. Using the third Kepler law we obtain a disc radius of
$\sim 3.1~10^{10}$~cm. The disc in DD~Cir is then large enough to reach the
3:1 resonance radius calculated from the equation (3.39) from Warner (1995);
$r_{\rm 3:1}\sim 3~10^{10}$~cm. The result is marginally in contrast with the
absence of superhumps. Taking the derived values for V5116~Sgr we get $r_{\rm
  3:1}\simeq 3.5~10^{10}$~cm as an upper limit of the disc radius taking
the absence of superhumps into consideration. The boundaries of the eclipse in
our light curve are not so clear as in the DD~Cir case, therefore not suitable
to estimate the disc radius.

Comparison to other systems follows the same way as in the case of IM~Nor and
DD~Cir. We took components masses from the Catalogue of Ritter and Kolb (2003)
and other masses are estimated from the orbital period (using Smith and
Dhillon 1998) and $t_{\rm 3}$ time (using Livio 1992). We rejected systems
with unknown or non measurable $t_{\rm 2}$ or $t_{\rm 3}$ time (peculiar light
curve, strong oscilations in the decline) without catalogue mass values (no
information on mass ratio, ex: RS~Car), magnetic novae (AM~Her systems without
disc, ex: V1500~Cyg), systems with orbital period $> 10$ h (Smith and Dhillon
fitting not adequate, systems with evolved secondary are then excluded too,
ex: DI~Lac, V841~Oph, V723~Cas, GK~Per), systems with insuficient photometric
observations to detailed period analysis (ex: V500~Aql, V446~Her, HZ~Pup,
DY~Pup, CT~Ser). The final list of systems is in Table~\ref{systems}.
\begin{figure}
\includegraphics[width=90mm]{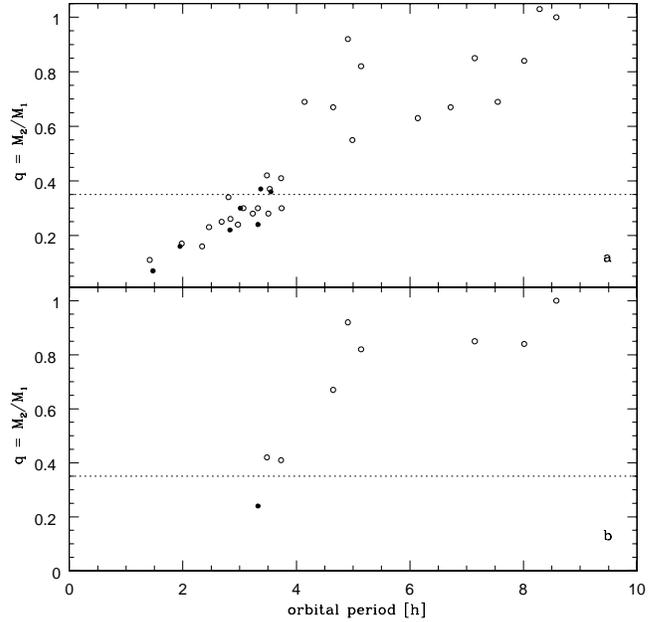}
\caption{Selected nova remnants in the mass ratio ($q$) -- orbital period
  ($P_{\rm orb}$) plane. Open circle -- non detected superhumps, filled circle
  -- detected superhumps. Panel a -- mass ratio from combined sourcees
  (catalogue and estimates from $P_{\rm orb}$, $t_{\rm 3}$ time), panel b --
  mass ratio only from catalogue values (Ritter and Kolb 2003). The dashed
  line is the 0.35 limit.}
\label{q_p}
\end{figure}

The results are depicted in Fig.~\ref{q_p}. The lower panel only shows the
systems with known catalogue mass values and the upper panel is after the
addition of estimated mass values in this work. Filled circles are systems
with detected superhumps. The critical mass ratio of 0.35 is also shown. The
results presented in the lower panel are as expected from the mass ratio 
limit for superhumps, but the statistical set is small. Adding other mass
estimates changes the situation. 7 systems below or close to the limit 0.35
show superhumps but 14 systems do not. The estimated mass ratio has a
mean difference of 0.2 from the catalogue values. The differences are
distributed randomly. Therefore there is no reason to suspect that all mass
ratios are systematically underestimated. The 7 systems with detected
superhumps are safely located in comparing to the 0.35 critical value as
expected by theory and are only a third of all systems falling below or close
to the mass ratio of 0.35. The majority of all systems occupy the orbital
period interval of 2 -- 4 hours. Following Patterson (2005) this interval has
a $\sim$ 40 -- 90 \% probability to observe superhumps. 7 systems versus 14
from our investigation yields only $\sim 30$ \%.

Superhumps are observed during the outbursts of short orbital period
dwarf novae (SU~UMa stars) where the disc reaches the 3:1 resonance
radius (superoutburst case). The disc during this active stage (outburst --
active and hot stage, quiescence -- not active and cold stage) reaches larger
radius that in quiescence (see Lasota 2001 for review). The irradiation of the
secondary by the central white dwarf in nova remnants can produce enhanced
mass transfer rate strong enough to keep the disc in this stable hot active
stage (as in the nova like stars) with the larger diameter (Hameury and Lasota
2005). In addition, superoutbursts in SU~UMa stars, when superhumps are
observed, can be explained by the irradiation and the enhanced mass transfer
(Smak 1995, 2004, Hameury et al. 2000, Schreiber et al. 2004). Therefore,
permanent superhumpers are possible after a nova outburst. A few possible
interpretation of the superhump lack can be: 1) not systematically applicable
tidal instability theory; tidal torques are probably not the main and only
condition of superhump existence as concluded by Hameury and Lasota (2005), 2)
disc radius not developed up to tidal radius, 3) superhumps present but not
detected, because of insufficient sets of data.
\section{Summary}

By using the maximum visual magnitude of $V = 7.15$ we measured the decay time
$t_{\rm 2} = 6.5 \pm 1.0$~d from the long-term light curve. This makes
V5116~Sgr a very fast nova. We estimated a visual absolute magnitude in
maximum of $M_{\rm Vmax} = -8.85 \pm 0.04$~mag. We found a period of
$0.1238 \pm 0.0001$~d ($2.9712 \pm 0.0024$~h) which we interpret as the
orbital period of the binary system. We propose that V5116~Sgr is a high
inclination system showing a strong irradiation effect of the secondary
star. No fully developed accretion disc up to the tidal radius is possible
with radius lower than $3.5~10^{10}$~cm.

Searching for superhumps in nova remnants using the limiting value for mass
ratio 0.35 and estimated mass values shows, that in the majority of systems
with expected superhumps this variability has not been found yet. Looking for
superhumps in these systems could be fruitful.

\section{Acknowledgments}

This work was supported by the Slovak Academy of Sciences Grant
No. 2/7011/7. We acknowledge with thanks the variable star observations from
the AAVSO International Database contributed by observers worldwide and used
in this research. We also acknowledge Michael Friedjung for valuable comments.

% \appendix

% \section{}
% \label{}

% Bibliographic references with the natbib package:
% Parenthetical: \citep{Bai92} produces (Bailyn 1992).
% Textual: \citet{Bai95} produces Bailyn et al. (1995).
% An affix and part of a reference:
%   \citep[e.g.][Ch. 2]{Bar76}
%   produces (e.g. Barnes et al. 1976, Ch. 2).

\newpage
\begin{table}[t]
\onecolumn
\caption{Selected nova systems. $P_{\rm orb}$ is the orbital period in hours,
  $M_{\rm 2}$ is the secondary mass in solar masses derived from Smith and
  Dhillon (1998) (from Ritter and kolb 2003 in parentheses), $t_{\rm 3}$ is
  the decline time of three magnitudes from maximum (value in parentheses is
  the value estimated from the $t_{\rm 2}$ time using $t_{\rm 3} \simeq
  2.75~t_{\rm 2}^{0.88}$ from Warner 1995), $M_{\rm 1}$ is the primary mass
  derived from Livio (1992) (from Ritter and kolb 2003 in parentheses), $q$ is
  the mass ratio, SH is the information about superhump detection (Y -- yes, N
  -- no) and Ref. is the reference.}
\begin{center}
\begin{tabular}{lccccccr}
\hline
\hline
 System & $P_{\rm orb}$ & $M_{\rm 2}$ & $t_{\rm 3}$ & $M_{\rm 1}$ & $q$ & SH
 & Ref.\\
  & [h] & [M$_{\rm \odot}$] & [d] & [M$_{\rm \odot}$] & & & \\
\hline
RW UMi & 1.418 & 0.07 & 140 & 0.64 & 0.11 & N & 19 \\
CP Pup & 1.474 & 0.08 & 8 & 1.19 & 0.07 & Y & 22 \\
V1974Cyg & 1.950 & 0.14 & 42 & 0.90 & 0.16 & Y & 9,26 \\
RS Car & 1.980 & 0.14 & (58) & 0.83 & 0.17 & N & 24 \\
DD Cir & 2.339 & 0.18 & (10) & 1.16 & 0.16 & N & 4 \\
IM Nor & 2.462 & 0.20 & 50 & 0.86 & 0.23 & N & 3,13 \\
QU Vul & 2.682 & 0.23 & 40 & 0.91 & 0.25 & N & 29 \\
V2214 Oph & 2.804 & 0.24 & 100 & 0.71 & 0.34 & N & 17 \\
V630 Sgr & 2.830 & 0.25 & 11 & 1.14 & 0.22 & Y & 23 \\
V351 Pup & 2.837 & 0.25 & (32) & 0.95 & 0.26 & N & 23 \\
V5116 Sgr & 2.971 & 0.26 & 20.2 & 1.04 & 0.25 & N & this paper \\
V4633 Sgr & 3.014 & 0.27 & 42 & 0.90 & 0.30 & Y & 14 \\
DN Gem & 3.068 & 0.28 & 37 & 0.92 & 0.30 & N & 28 \\
V1494 Aql & 3.231 & 0.30 & 16 & 1.08 & 0.28 & N & 8,30,31 \\
V1668 Cyg & 3.322 & 0.31 & 23 & 1.02 & 0.30 & N & 32 \\
V603 Aql & 3.324 & 0.31(0.29) & 8 & 1.19(1.2) & 0.26(0.24) & Y & 21 \\
RR Cha & 3.370 & 0.31 & 58 & 0.83 & 0.37 & Y & 24,1 \\
RR Pic & 3.481 & 0.33(0.4) & 150 & 0.62(0.95) & 0.53(0.42) & N & 33,34 \\
V382 Vel & 3.508 & 0.33 & 9 & 1.17 & 0.28 & N & 2,12 \\
V533 Her & 3.530 & 0.33 & 44 & 0.89 & 0.37 & N & 35,42 \\
V2574 Oph & 3.550 & 0.34 & (33) & 0.95 & 0.36 & Y & 15 \\
OY Ara & 3.731 & 0.36(0.34) & ... & (0.82) & (0.41) & N & 36 \\
V1493 Aql & 3.740 & 0.36 & 7 & 1.21 & 0.30 & N & 8,11 \\
V849 Oph & 4.146 & 0.41  & 175 & 0.59 & 0.69 & N & 37 \\
DQ Her & 4.647 & 0.48(0.40) & 94 & 0.72(0.60) & 0.67(0.67) & N & 38 \\
T Aur & 4.906 & 0.51(0.63) & 100 & 0.71(0.68) & 0.72(0.92) & N & 41 \\
V4745 Sgr & 4.987 & 0.52 & 32.8 & 0.95 & 0.55 & N & 7,10 \\
HR Del & 5.140 & 0.54(0.55) & 230 & 0.54(0.67) & 1.00(0.82) & N & 43 \\
V1425 Aql & 6.139 & 0.57 & 39 & 0.91 & 0.63 & N & 8,27 \\
V4743 Sgr & 6.718 & 0.74 & 15 & 1.10 & 0.67 & N & 16,39 \\
BY Cir & 6.760 & 0.74 & 157 & 0.62 & 1.19 & N & 4 \\
V2540 Oph & 6.835 & 0.75 & 213 & 0.55 & 1.36 & N & 18 \\
V838 Her & 7.143 & 0.79(0.74) & 3.2 & 1.29(0.87) & 0.61(0.85) & N & 44 \\
V2275 Cyg & 7.548 & 0.84 & 7 & 1.21 & 0.69 & N & 20,40 \\
BT Mon & 8.012 & 0.90(0.87) & (213) & 0.55(1.04) & 1.64(0.84) & N & 45 \\
V368 Aql & 8.285 & 0.93 & 42 & 0.90 & 1.03 & N & 46 \\
QZ Aur & 8.580 & 0.97(1.05) & 27 & 0.99(1.05) & 0.98(1.00) & N & 47 \\
\hline
\hline
\end{tabular}
\end{center}
Referencees:\\
1 - Hoffmeister (1959), 2 - Balman et al. (2006), 3 - Woudt and Warner
(2003b), 4 - Woudt and Warner (2003c), 5 - Lynch et al. (2004), 6 - Barlow et
al. (1981), 7 - Cs\'ak et al. (2005), 8 - Arkhipova et al. (2002), 9 -
Skillman et al. (1997), 10 - Dobrotka et al. (2006a), 11 - Dobrotka et
al. (2006b), 12 - Della Valle et al. (2002), 13 - Kato et al. (2002a), 14 -
Lipkin et al. (2001), 15 - Kang et al. (2006a), 16 - Kang et al. (2006b), 17 -
Baptista et al. (1993), 18 - Ak et al. (2005), 19 - Retter and Lipkin (2001),
20 - Balman et al. (2005), 21 - Retter and Naylor (2000), 22 - Patterson and
Warner (1998), 23 - Woudt and Warner (2001), 24 - Woudt and Warner (2002), 25
- Matsumoto et al. (2001), 26 - Retter et al. (1997), 27 - Retter and
Leibowitz (1998), 28 - Retter et al. (1999), 29 - Shafter and Misselt (1995),
30 - Kato et al. (2004), 31 - Barsukova and Goranskii (2003), 32 - Kaluzny
(1990), 33 - Vogt (1975), 34 - Warner (1986), 35 - Patterson (1979a), 36 -
Zhao and McClintock (1997), 37 - Shafter et al. (1993), 38 - Walker (1961), 39
- Kato et al. (2002b), 40 - Kiss et al. (2002), 41 - Walker (1963), 42 -
Thorstensen and Cynthia (2000), 43 - Kohoutek and Pauls (1980), 44 - Leibowitz
et al. (1992), 45 - Robinson et al. (1982), 46 - Diaz and Bruch (1994), 47 -
Campbell and Shafter (1995)\\
Aditional sourcees for orbital period, $t_{\rm 2}$/$t_{\rm 3}$ time and
catalogue masses:\\
Duerbeck (1987), Warner (1986, 1995, 2002), Diaz and Bruch (1997), Ritter and
Kolb (2003)
\label{systems}
\twocolumn
\end{table}

\end{document}